\documentclass[preprint,showpacs,preprintnumbers,amsmath,amssymb,superscriptaddress]{revtex4}
\usepackage{amssymb}
\usepackage{graphicx,color}
\usepackage{bm} 

\usepackage{graphicx} \usepackage{dcolumn} \usepackage{bm}
\newcommand{\beq}{\begin{equation}}\newcommand{\eeq}[1]{\label{#1}
\end{equation}}\newcommand{\beqar}{\begin{eqnarray}}\newcommand{\eeqar}[1]
{\label{#1}
\end{eqnarray}}\newcommand{\bmath}{\begin{displaymath}}\newcommand{\emath}{\end{displaymath}}\newcommand{\bitem}{\begin{itemize}}\newcommand{\eitem}{\end{itemize}}
 
\begin{document}

\title{\Large \bf Hyperon/meson ratios   
     in rare high-multiplicity $pp$ collisions at
     energies available at the
     Large Hadron Collider, and potential signatures 
     for mini-quark-gluon plasma formation}

\newcommand{\mcgill}{McGill University, Montreal, Canada, H3A 2T8}

\newcommand{\columbia}{Columbia University, New York, N.Y. 10027}

\affiliation{\mcgill}
\affiliation{\columbia}

\author{~V.~Topor~Pop} \affiliation{\mcgill}
\author{~M.~Gyulassy} \affiliation{\columbia} 
\author{~J.~Barrette} \affiliation{\mcgill}
\author{~C.~Gale} \affiliation{\mcgill}
\author{~A.~Warburton} \affiliation{\mcgill}

\date{August 23, 2012}

\begin{abstract}

We use the framework of the HIJING/B\=B v2.0 model to simulate 
high-multiplicity (HM) $p+p$ collision events at the 
Large Hadron Collider (LHC) to study 
observables sensitive to possible collective phenomena, such as strong 
longitudinal color fields (SLCF) modeled by an enhanced string 
tension ($\kappa$).
We focus on the hyperon/meson yield ratios at
center-of-mass (c.m.) energy $\sqrt{s}$ = 7 TeV, in the transverse momentum 
region, $1 < p_T < 4 $ GeV/{\it c}. 
For minimum bias events these ratios are well described assuming 
an energy dependence  
$\kappa = \kappa(s)= \kappa_{0} (s/s_{0})^{0.04}\,\,{\rm GeV/fm}$\\
~($\kappa_{0}$= 1~GeV/fm),
giving a value $\kappa = 2$ GeV/fm at $\sqrt{s}$ = 7 TeV.
We compare minimum bias (MB) events
to simulated HM events assuming that $\kappa(MB)=2$ GeV/fm 
could grow to an extreme value of $\kappa(HM)=5$ GeV/fm 
that saturates the strangeness suppression factor.
With this assumption the model predicts a very strong enhancement 
of (multi)strange baryon/meson ratios in HM events. If observed, 
such an enhancement could be also interpreted as a possible signature 
for formation in HM $p+p$ collision events 
of a deconfined but out of local thermal equilibrium 
{\em mini quark-gluon plasma} (mQGP).

\end{abstract}

\pacs{12.38.Mh, 24.85.+p, 25.40Ve, 25.75.-q}

\maketitle




\section{Introduction}

Charged particle multiplicities measured in high-multiplicity (HM)
$p+p$ collisions at CERN Large Hadron Collider (LHC) energies   
reach values that are of the same
order as those measured in heavy-ion collisions at lower energies
({\it e.g.}, well above those observed at RHIC for Cu + Cu 
collisions at $\sqrt{s_{\rm NN}}$ = 200 GeV \cite{Alver:2010ck}).
The Bjorken energy density relation \cite{Bjorken:1982qr} connects   
high multiplicity events with high energy density.
Within that approach 
at the LHC, $p+p$ collisions could reach an energy density of 5-10 GeV/fm$^3$, 
comparable to those in $A+A$ collisions at RHIC \cite{Adler:2004zn}.
It is, therefore, a valid question whether $p+p$ collisions also exhibit 
any behavior of the kind observed in heavy-ion collisions
\cite{bjorken_1982,Abreu:2007kv,Cunqueiro:2008uu,Werner:2010ny,Kisiel:2010xy,Bozek:2010pb,Pirner:2011ab,Braun:2012kn}.
Bjorken first suggested the idea of possible deconfinement in $p+p$ collisions
\cite{bjorken_1982}. It has also been suggested by Van Hove 
\cite{VanHove:1982vk} and recently by Campanini \cite{Campanini:2011bj}
that an anomalous behavior of average
transverse momentum ($<p_T>$) as a function of multiplicity could be
a signal for the occurrence of a phase transition in hadronic matter,
{\it i.e.}, formation of a {\em mini quark-gluon plasma} (mQGP).
The hadronic interaction model EPOS 
(Partons Off-shell remnants and Splitting of 
parton ladders), 
has also been used to describe the production of mQGP features in high
energy density $p+p$ collisions \cite{Liu:2011np,Liu:2011dk}. 

Another indication of collective phenomena 
might be the observed long-range,
near-side angular correlation (ridge) 
in HM $p+p$ collisions at center-of-mass energy 
$\sqrt{s}$ = 7 TeV for charged particle multiplicities 
well above the mean multiplicity.
CMS~\cite{Khachatryan:2010gv,Li:2011mp,Velicanu:2011zz}  
and ATLAS \cite{atlas_conf2011} constructed a two-particle 
correlation function 
and measured its value for different $\Delta \eta$ and $\Delta \Phi$
angular separations.
When looking at particles in a specific range of $p_T$ 
and high multiplicity ($1 < p_T < 3$ GeV/{\it c} and $N_{\rm ch} > 110$),
a clear ridge-like structure emerges at  $\Delta \Phi$ $\approx$ 0  and 
$2 < |\Delta \eta| < 4.8$,
that is not reproduced by existing Monte Carlo (MC) 
event generators~\cite{Khachatryan:2010gv}.
The origin of this unexpected {\em ridge-like} structure 
found in the two-particle
correlation analysis, albeit
attracting much theoretical attention, is still under debate
\cite{Werner:2010ss,Azarkin:2011xh,Hwa:2010fj,Dumitru:2010iy,Trainor:2010uk,Shuryak:2010wp}.

Identified particle production  
has been studied in detail by the ATLAS 
\cite{Aad:2011hd}, ALICE \cite{Aamodt:2011zza,Aamodt:2011zj,Chojnacki:2011pv,Floris:2011ru,Chinellato:2011yn,Abelev:2012jp}
and CMS \cite{Khachatryan:2011tm,Rougny:2012yx,cms_id_march2012} 
collaborations in $p+p$ collisions at the LHC. 
Meson ($\pi$, $K$, $K_S^0$) and baryon ($p$, $\Lambda$, $\Xi^-$, $\Omega$)
yields, rapidity and
multiplicity distributions have been measured with different event
selections [minimum bias, inelastic (INEL) or non-single
diffractive events (NSD)].
For minimum bias event selection, different {\small PYTHIA} parameter
sets~\cite{Sjostrand:2007gs,Sjostrand:2006za,Skands:2010ak,Field:2010bc}
 have difficulty reproducing 
(multi)strange particle production, predicting 
too few strange particles and harder $p_T$ spectra,
the differences with data increasing with the mass of the strange
particle \cite{Rougny:2012yx}. 
Up to now, none of the MC event generators is able 
to describe the softer $p_T$ and the huge rise of particle production 
with energy. This has led to a concerted effort to improve  
the available MC generators. 

In a string fragmentation phenomenology, it has been proposed
that the observed strong enhancement of strange particle
production in nuclear collisions
could be naturally explained via strong longitudinal 
color field effects (SLCF)~\cite{Gyulassy:1986jq}.
Recently, an extension
of Color Glass Condensate (CGC) theory has proposed a more detailed
dynamical ``GLASMA'' model \cite{larry_2009,mclerran_08} of color ropes.
In the string models, strong longitudinal fields 
(flux tubes, effective strings) decay into new ones by 
 quark anti-quark ($q\bar{q}$ ) or 
diquark anti-diquark (qq-$\overline{\rm qq}$) pair 
production and subsequently hadronize to produce the 
observed hadrons. Due to confinement, the color 
of these strings is restricted to a small area in transverse space.
With increasing energy of the colliding particles, the number of
strings grows and they start to overlap, forming clusters 
This can introduce a possible dependence of particle production on the
energy density \cite{Braun:2012kn}.

We have studied~\cite{ToporPop:2010qz} the effect of  strong 
longitudinal color fields (SLCF)
in $p+p$ collisions up to LHC energies in the framework of
the {\small HIJING/B\=B } v2.0 model,  
which combines (collinear factorized) pQCD
multiple minijet production with soft longitudinal string excitation and
hadronization. The default vacuum string tension, $\kappa_0$ = 1 GeV/fm, 
is replaced by an effective energy dependent string tension, 
$\kappa(s) = \kappa_0 (s/s_0)^{0.06}$ GeV/fm that increases monotonically with 
center-of-mass energy. The exponent $\lambda=0.06$ is found
to succeed at describing well the energy
dependence of multiparticle observables for RHIC, Tevatron, as
well as LHC data \cite{ToporPop:2010qz}. 
In the {\small HIJING/B\=B} v2.0 model the rapid growth
of $dN_{ch}/d\eta$ at mid-rapidity with energy is due to the interplay 
of copious minijet
production with increasing strong color field contributions.
However, the large (strange)baryon-to-meson ratios recently measured at 
LHC energies, especially at $\sqrt{s}$ = 7 TeV, are not well 
described using this set of parameters. 

In this work we will address this question and in addition we will
discuss a possible dependence of the strength of strong color field 
on the event multiplicity. We will show that the model predicts 
a very strong enhancement of (multi)strange baryon-to-meson ratios 
in HM events. If observed, this could be interpreted as a possible
signature for formation of a deconfined but out of local thermal
equilibrium mini quark-gluon plasma.

\section{The effective string tension: an infrared sensitive dynamical variable}

For a uniform chromo-electric flux tube with field ({\it E}), 
the pair production rate \cite{Gyulassy:1986jq,cohen_jul08}  
per unit volume for a (light)heavy quark ($Q$) is given by
\begin{equation}
\Gamma =\frac{\kappa^2}{4 \pi^3} 
{\text {exp}}\left(-\frac{\pi\,m_{Q}^2}{\kappa}\right),
\label{eq:eq1}
\end{equation}
where $Q={\rm qq}$ (diquark), $s$ (strange), $c$ (charm) or $b$
(bottom). The {\em current 
quark masses} are $m_{\rm qq}$ = 0.45 GeV \cite{ripka:2005},  
$m_s=0.12$ GeV, $m_{\rm c}=1.27$ GeV, and $m_b=4.16$ GeV \cite{pdg:2010}.
The {\em constituent quark masses} of light non-strange quarks 
are $M_{u,d}$ = 0.23 GeV, of the strange quark is $M_s$=0.35 GeV 
\cite{armesto2001}, and of the diquark is $M_{\rm qq}=0.55 \pm 0.05$ GeV 
\cite{ripka:2005}.

An enhanced rate for spontaneous pair production is naturally
associated with 
``{\em strong chromo-electric fields}'', such 
that $\kappa/m_{\rm Q}^2\,\,>$ 1 {\em at least some of the time}.
In a strong longitudinal color electric field, 
the heavier flavor suppression factor 
$\gamma_{Q\bar{Q}}$ varies with string tension 
via the well known Schwinger formula \cite{schwinger}, 

\begin{equation}
\gamma_{Q\bar{Q}} = \frac{\Gamma_{Q\bar{Q}}}{\Gamma_{q\bar{q}}} =
{\text {exp}} \left(-\frac{\pi(M_{Q}^2-m_q^2)}{\kappa_0} \right)
 < 1
\label{eq:eq2}
\end{equation}
for $Q = {\rm qq}$, $s$, $c$ or $b$ and $q = u$, $d$.

In the model calculations, we assume the following effective masses: 
$M^{\rm eff}_{qq}$ = 0.5 GeV, 
$M^{\rm eff}_{s}$ = 0.28 GeV, and $M^{\rm eff}_{c}$ = 1.30 GeV.
Therefore, the above formula implies a 
suppression of heavier quark production according to
$u$ : $d$ : ${\rm qq}$ : $s$ : $c$ $\approx$ 1 : 1 : 0.02 : 0.3 : 10$^{-11}$ 
for the vacuum string tension $\kappa_0$ = 1 GeV/fm.
For a color rope (or cluster), on the other hand,
if the {\em average string tension} value $\kappa$ 
increases, the suppression factors $\gamma_{Q\bar{Q}}$ increase
(i.e., this implies a higher rate of $Q\bar{Q}$ pair production).

Using the {\small HIJING/B\=B} v2.0 model, 
we have shown that it is important to consider that high energy 
$p+p$ collisions can have a substantial contribution 
from SLCF effects~\cite{ToporPop:2010qz}.
In the model phenomenology, the degree of collectivity is described
by the overlap of individual strings (clusters), quantified 
by the infrared sensitive variable, the string tension ($\kappa$).
A reduction mechanism of strange 
quark suppression was introduced 
by assuming that the effective string tension increased with
increasing reaction energy according to a power law: 

\begin{equation}
\kappa = \kappa(s)= \kappa_{0} \,\,(s/s_{0})^{0.06}\,\,{\rm GeV/fm},
\label{eq:eq3}
\end{equation} 

where $\kappa_{0}$ = 1 GeV/fm is the vacuum string tension value and 
$s_{0}$ = 1 GeV$^2$ is a scale factor.
In addition to describing well the energy dependence of charged
particle density at mid-rapidity from SPS to LHC energies,
we have shown that this dynamical mechanism improves the 
description of the strange meson/hyperon data at Tevatron and LHC energies
\cite{ToporPop:2010qz,Pop:2009sd,ToporPop:2011wk}.
However, using a set of parameters corresponding to those obtained 
from Eq.~\ref{eq:eq3}, results in an over-prediction
of the recently measured yields   
\cite{Chojnacki:2011pv,Floris:2011ru,Chinellato:2011yn,Abelev:2012jp}
 of $\Xi$ particles (by up to a factor of two) and of $\Omega$ particles 
(by up to a factor of four).

A possible reason for this could be a too strong  
energy dependence used for the mean string tension values
(see Eq.~\ref{eq:eq3}).  
Therefore, here we consider a weaker energy dependence of the form, 
\begin{equation}
\kappa = \kappa(s)= \kappa_{0} (s/s_{0})^{0.04}{\rm GeV/fm},
\label{eq:eq4}
\end{equation} 
This leads to a value for the mean string tension of   
$\kappa = \kappa (s) = 1.7 $~GeV/fm at 0.9 TeV and 
$\kappa = \kappa (s) = 2 $~GeV/fm at 7 TeV.
The results obtained using this new parametrization will be discussed
in Section III.

Because the threshold for strange quark production in a 
deconfined phase (or in a mQGP)
is much smaller than in a hadron gas, a larger enhancement in strange 
particle production has been suggested as an indication of possible mQGP
formation \cite{muller:1986,Muller:2011tu,Muller:2012zq}. 
Equation~\ref{eq:eq2}, which describes strangeness suppression factors 
  $\gamma_{s\bar{s}}$, shows that an increased value $\kappa$ up to 
$\kappa \approx 5$ GeV/fm
leads to a saturation of suppression factors, $\gamma_{s\bar{s}}$. 
For example, in $p+p$ collisions at 7 TeV 
the suppression factor $\gamma_{s\bar{s}}$ increases from
0.4 (corresponding to $\kappa_0 = 1$ GeV/fm) to approximately 0.85 
(corresponding to $\kappa = 5$ GeV/fm), and has only 
a modest further increase up to 0.89 for $\kappa = 8-10$ GeV/fm.

Therefore, if we assume a different energy density
in MB and HM $p+p$ collision events 
(corresponding also to a different type of ropes formation) 
could lead to different effective $\kappa$ values.
 The $\kappa$ value obtained from Eq.~\ref{eq:eq4} 
$\kappa$= $\kappa(s) \approx 2 \kappa_0$
can be associated with MB events, since it is deduced from inclusive
data. For HM events we will assume an extreme value of 
$\kappa \approx 5 \kappa_0$ GeV/fm, 
which corresponds to a saturation of strangeness suppression factors. 
To better characterize the sensitivity to the parameter $\kappa$, we
also present the result corresponding to an intermediate value 
$\kappa$= $\kappa(s) \approx 3 \kappa_0$.

Note that a value $\kappa \approx 5 \kappa_0$ GeV/fm is also supported by 
recent calculations at finite temperature ($T$) 
of potentials associated with a $q\bar{q}$ pair separated by a
distance $r$ \cite{Liao:2008vj}.
The finite temperature ($T$) form of the $q\bar{q}$ potential
has been calculated by means of lattice QCD \cite{Kaczmarek:2005ui}.
At finite temperature, there are two potentials associated
with a $q\bar{q}$ pair separated by a distance $r$:
the free energy $F(T,r)$ and internal energy $V(T,r)$.
The free and internal energies actually correspond to slow and fast
(relative) motion of the charges, respectively \cite{Liao:2008vj}.
Infrared sensitive variables such as  string tension
are very helpful to identify specific
degrees of freedom of the plasma. 
Since the confinement of color in non-Abelian theories is due to
the magnetic degree of freedom, the magnetic component is 
expected to be present in
the plasma as well \cite{Chernodub:2006gu}.
In the presence of the {\em chromo-magnetic scenario} it was shown that 
the effective string tension of the free energy $\kappa$ = $\kappa_F$ 
decreases with $T$, to near zero at critical temperature ($T_c$)
\cite{Liao:2008vj}. 
In contrast, the effective string tension of the internal energy
(corresponding to a fast relative motion of the charges)
$\kappa$ = $\kappa_V$ remains nonzero below about $T = 1.3 \,T_c$
with a peak value at $T_c$ about 5 times the vacuum tension
$\kappa_{0} $ ($\kappa_V$ = 5 $\kappa_{0} $ = 5 GeV/fm)
\cite{Liao:2008vj}.

\section{ Results and Discussions}

\subsection{Charged hadron pseudorapidity and transverse-momentum spectra}

All details of the {\small HIJING/B\=B} v2.0 model are extensively 
discussed in the literature
\cite{ToporPop:2010qz,Pop:2009sd,ToporPop:2011wk}.
Here we focus our analysis at two energies of interest: 
$\sqrt{s}= 0.9$ TeV and $\sqrt{s}= 7$ TeV, where data for
 charged particles 
\cite{Khachatryan:2010xs,Khachatryan:2010us,Aamodt:2010my,Otwinowski:2011gq,Aamodt:2010pp,Aamodt:2010ft,Aad:2010rd,Aspell:2012ux}
and identified particles
\cite{Aamodt:2011zza,Aamodt:2011zj,Chojnacki:2011pv,Floris:2011ru,Chinellato:2011yn,Abelev:2012jp,Khachatryan:2011tm,Rougny:2012yx,cms_id_march2012}  
have been reported for MB events.
Except for the new energy dependence of the string tension 
(Eq.~\ref{eq:eq4}), all the other parameters of the models are as in 
Ref.~\cite{ToporPop:2010qz}.
Note that this modification also leads to a 
relatively small increase (approximately 10-12\%)
for the predicted density of charged particles at mid-rapidity, 
in comparison with previous results reported in Ref.~\cite{ToporPop:2010qz} 
in the entire energy region of interest. 
This increase is now also supported by the new experimental data
with LHC energy $\sqrt{s} = 7$ TeV
\cite{Khachatryan:2010xs,Khachatryan:2010us,Aamodt:2010my,Otwinowski:2011gq,Aamodt:2010pp,Aamodt:2010ft,Aad:2010rd,Aspell:2012ux}.

\begin{figure} [h!]
\centering
\includegraphics[width=0.9\linewidth,height=7.0cm]{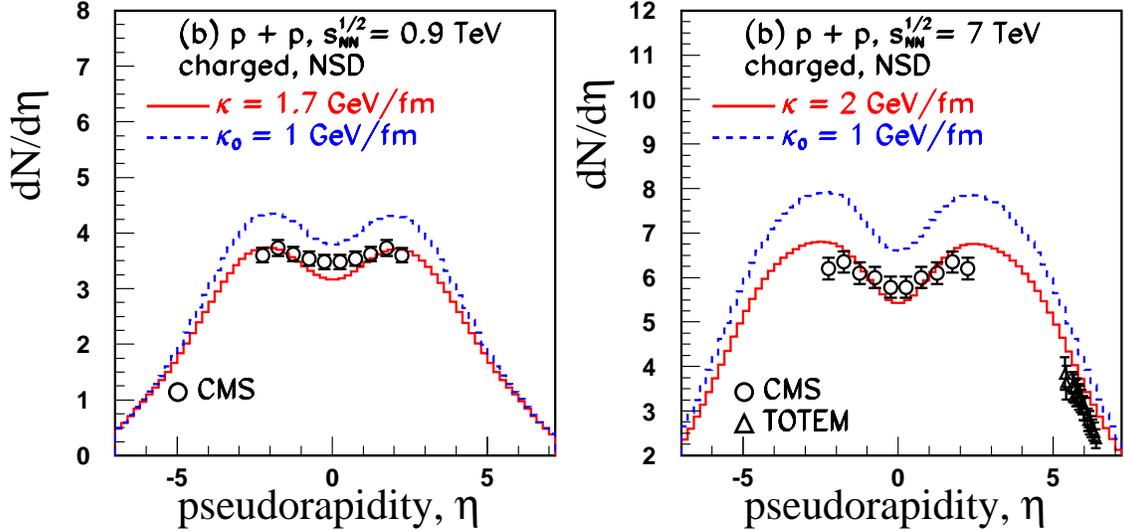}
\vskip 0.5cm\caption[NSD dn/deta cahrged hadrons]
{\small (Color online) Comparison of {\small HIJING/B\=B v2.0} 
predictions for charged particle pseudorapidity 
distributions at $\sqrt{s} = 0.9$ TeV (left panel)
and $\sqrt{s} = 7$ TeV (right panel) 
for non-single-diffractive (NSD) $p+p$ collisions.
The solid and dashed histograms are the results
with and without SCF, respectively.
The data are from Refs.~\cite{Khachatryan:2010xs,Khachatryan:2010us}
(CMS) and from Ref.~\cite{Aspell:2012ux} (TOTEM). 
Only statistical error bars are shown. 
\label{fig:fig1}
}

\end{figure}

Charged hadron multiplicity measurements are the first results 
of the LHC physics program. 
The new data on charged particle pseudorapidity distributions
\cite{Khachatryan:2010xs,Khachatryan:2010us}, 
over a limited $\eta$ range 
for non single diffractive interactions (NSD), 
are compared to model calculations  in Fig.~\ref{fig:fig1}. 
The data show a sizeable increase of the central pseudorapidity
density with c.m.s. energy.
As the colliding energy increases, the rate of
multiple parton interactions (MPI)
also increases, producing a rise in the central multiplicity.
The increase with energy in our phenomenology
is due to the interplay of the increased mini-jet production
in high colliding energy with SLCF effects.
A scenario with SLCF effects (solid
histograms) reproduces well the measured multiplicity distributions.
Without SLCF effects ({\it i.e.}, $\kappa= \kappa_0 = 1$ GeV/fm )
the model strongly overestimates the central
charged particle density (dashed histograms). 
Data over a larger rapidity range are needed 
to determine the shape of the falling
density in the fragmentation region. For completeness, 
the new TOTEM data \cite{Aspell:2012ux} 
at forward pseudorapidity are also included. 


\begin{figure} [h!]

\centering

\includegraphics[width=0.9\linewidth,height=7.0cm]{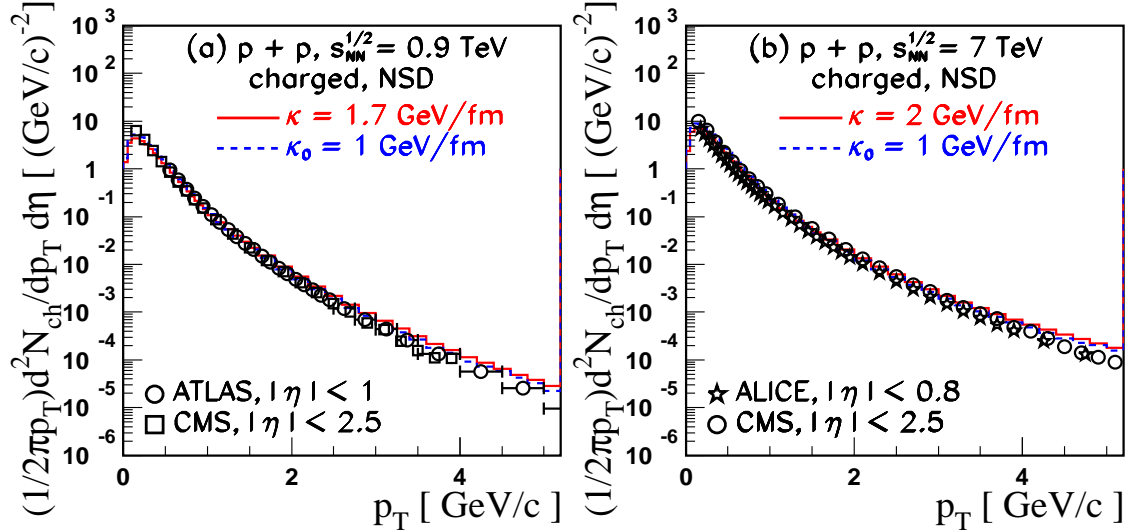}
\vskip 0.5cm\caption[transverse momentum distribution 0.9-7 TeV]
{\small (Color online) Comparison with data 
of {\small HIJING/B\=B v2.0} predictions of
charged-hadron transverse momentum distributions ($|\eta| < 0.8$) 
at LHC energies.
The calculated spectra include the combined effects of SLCF and J\=J
loops. The solid and dashed histograms have the same meaning as 
in Fig.~\ref{fig:fig1}.
The data are from Refs.~\cite{Khachatryan:2010xs,Khachatryan:2010us} (CMS),
\cite{Otwinowski:2011gq} (ALICE preliminary), \cite{Aad:2010rd} (ATLAS).
Statistical error bars on the data points are smaller than the markers. 
\label{fig:fig2}
}

\end{figure}

The measured transverse momentum distributions for NSD events 
 at $\sqrt{s}$  = 0.9 TeV  
and $\sqrt{s}$ = 7 TeV are shown in Fig.~\ref{fig:fig2}
in  the range $0 < p_T < 5 $ GeV/c, where
both hard and soft processes are expected to contribute.
The data of CMS \cite{Khachatryan:2010xs,Khachatryan:2010us} 
and ATLAS \cite{Aad:2010rd} are measured in larger pseudorapidity
intervals $|\eta| < 2.5 $ and $|\eta| < 1 $, respectively. 
In contrast, ALICE measurements 
\cite{Otwinowski:2011gq} are in a very central region $|\eta| < 0.8$. 
The calculations were performed using ALICE acceptance but, as can
inferred from the data shown in Fig.~\ref{fig:fig1},
the difference in pseudorapidity range has a negligible effect  
on the $p_T$ spectra.
The model calculations including SLCF effects
give a good description of the spectral shape at low $p_T$ ($p_T < 4$ GeV/c)
for both energies. At high $p_T$ ($p_T > 5$ GeV/c) the calculations
lead to a somewhat harder spectrum than that observed. 
In our phenomenology this could indicate that jet quenching, {\it i.e.},
suppression of high $p_T$ particles like that observed at RHIC energies
in nucleus-nucleus collisions, could also appear in $p+p$ collisions,
particularly for events with large multiplicity~\cite{ToporPop:2010qz}.

\subsection{Minimum Bias events. Identified Particle Spectra and Ratios.}

The $pp$ single particle inclusive $p_T$ spectra measurements
are important for understanding collision dynamics, since
the various particles show different systematic behavior,
as observed at RHIC energy \cite{Tannenbaum:2010ab}. 
Detailed theoretical predictions for single inclusive hadron
production (including hyperons) are discussed in this section.
Baryon-to-meson ratios 
are experimental observables that can be used at the LHC   
to investigate multi-parton interactions 
and help understanding of the underlying physics
\cite{Hippolyte:2009xz,Ricaud:2010ay}.
Unexpectedly high ratios observed in $A+A$ collisions at RHIC energies 
have been discussed in terms of recombination and coalescence mechanisms
\cite{Fries:2003kq,Greco:2003mm,Hwa:2002tu}. In $p+p$ collisions, however,
a coalescence/hadronization scenario  is not
favored due to low phase space density in the final state.
The {\small HIJING/B\=B} model \cite{ToporPop:2010qz,ToporPop:2011wk},
with SLCF effects and junction-anti-junction ($J\bar{J}$) loops included,
provides an alternative dynamical
explanation of the heavy-ion data at RHIC energies.
We have shown that the model also
predicts an increasing yield of (multi)strange particles, thereby 
better describing the experimental data in $A+A$ and $p+p$ collisions.


\begin{figure} [h!]
\centering
\includegraphics[width=0.9\linewidth,height=7.0cm]{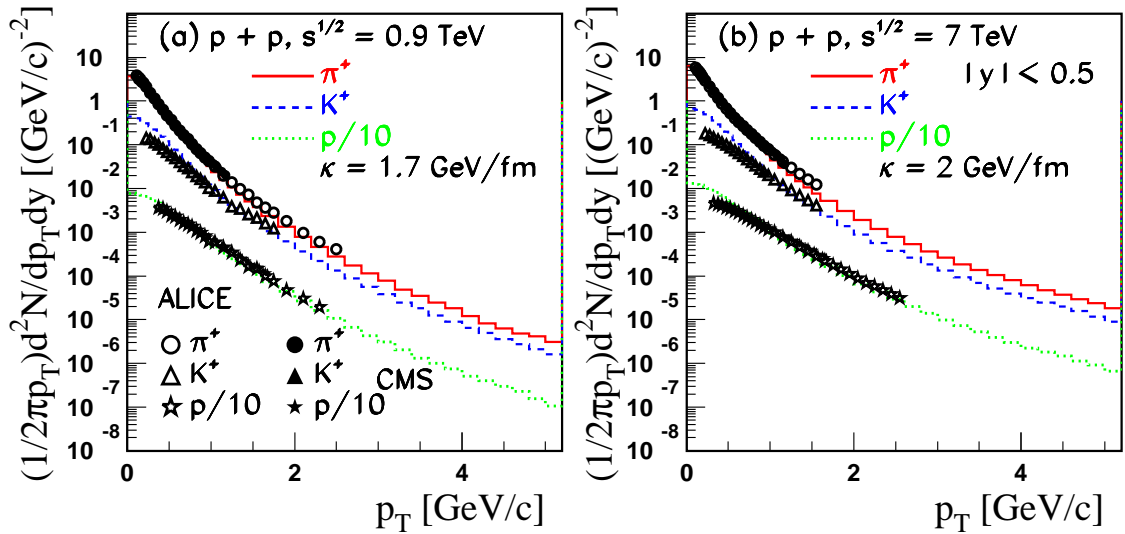}
\vskip 0.5cm\caption[strange baryon/mesons at 1.8 TeV -CDF data] 
{\small (Color online) {\small HIJING/B\=B v2.0}
  predictions of transverse momentum spectra at mid-rapidity ($|y|<0.5$)
 for mesons ($\pi,K$) and baryons ($p$)
 at $\sqrt{s}$ = 0.9 TeV (left panel) and at 
 $\sqrt{s}$ = 7 TeV (right panel) are compared to data.
The data (open symbols, left panel) are from Ref.~\cite{Aamodt:2011zj}
(ALICE). The preliminary data (open symbols, right panel) are
from Ref.~\cite{Chojnacki:2011pv} (ALICE).
Preliminary data (closed symbols) are from
CMS collaboration and are plotted using a normalization factor 0.78 
discussed in Ref.~\cite{cms_id_march2012}. 
Statistical error bars on the data points are smaller than the markers.
For clarity the data and theoretical calculations for the proton results  
are divided by a factor of ten.
 \label{fig:fig3}
}
\end{figure}

Figure~\ref{fig:fig3} compares the ALICE results to the predicted  
mid-rapidity spectra for positive pions (solid histograms), 
kaons (dashed histograms), and protons (dotted histograms)
in minimum bias $p+p$ collisions.    
There is  agreement in the $p_T$ region of interest 
$1 < p_T < 4$ GeV/{\it c} at both energies.
The over-prediction for proton and kaon production below $p_T = 1 $ GeV/{\it c}
is consistent with possible presence of radial flow,
which seems to be larger at 7 TeV than at 0.9 TeV.  
The radial flow could appear  
as a consequence of a hydrodynamic type evolution with flux tube
initial conditions \cite{Werner:2010ny}, not embedded in our model.     

\begin{figure} [h!]
\centering
\includegraphics[width=0.9\linewidth,height=7.0cm]{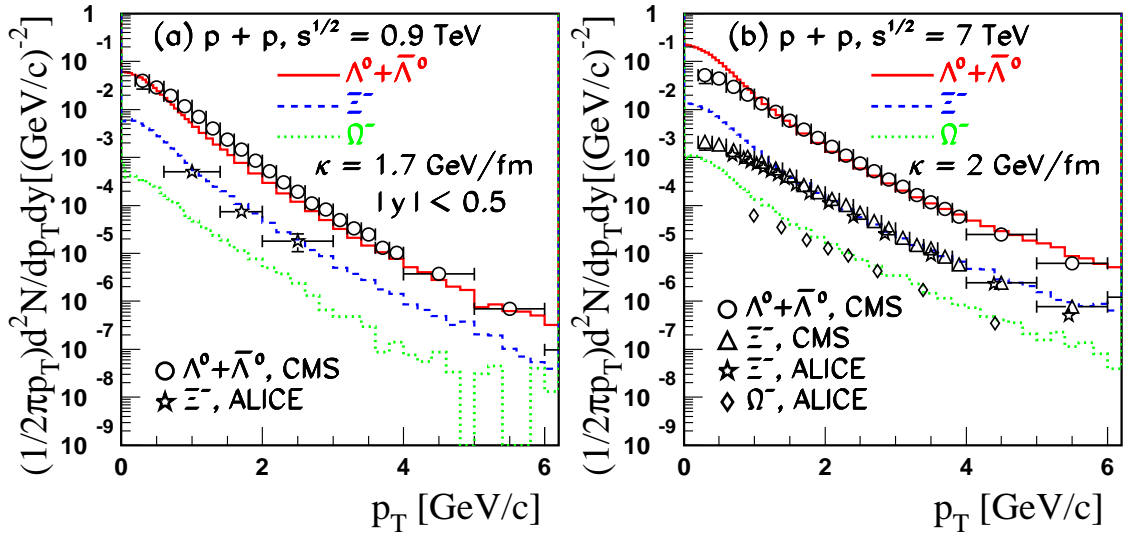}
\vskip 0.5cm\caption[strange baryons transverse spectra] 
{\small (Color online) {\small HIJING/B\=B v2.0}
  predictions of transverse momentum spectra at mid-rapidity ($|y|<0.5$)
 for (multi)strange hyperons ($\Lambda$, $\Xi$, $\Omega$)  
 at $\sqrt{s}$ = 0.9 TeV (left panel) and at 
 $\sqrt{s}$ = 7 TeV (right panel) are compared to data.
The data are from the ALICE ~\cite{Aamodt:2011zza,Abelev:2012jp}
and CMS collaborations ~\cite{Khachatryan:2011tm}.
Error bars include only the statistical uncertainties.  
\label{fig:fig4}
}
\end{figure}

We extend our analysis to the production of (multi)strange baryons. 
In Fig.~\ref{fig:fig4} we show the 
{\small HIJING/B\=B} v2.0 model predictions 
of $p_T$ spectra at mid-rapidity ($|y| < 0.5$) for  
$\Lambda$ (solid histograms), $\Xi^-$ (dashed histograms)
and $\Omega^-$ (dotted histograms) baryons at 
0.9 TeV and 7 TeV. For (multi)strange particles,
the data indicate a stronger radial flow at 7 TeV than at 0.9 TeV.
For the $p_T$ region of interest, $ 1< p_T < 4$ GeV/{\it c}, 
the model results are in agreement with data at both energies.

\begin{figure} [h!]
\centering
\includegraphics[width=0.9\linewidth,height=7.0cm]{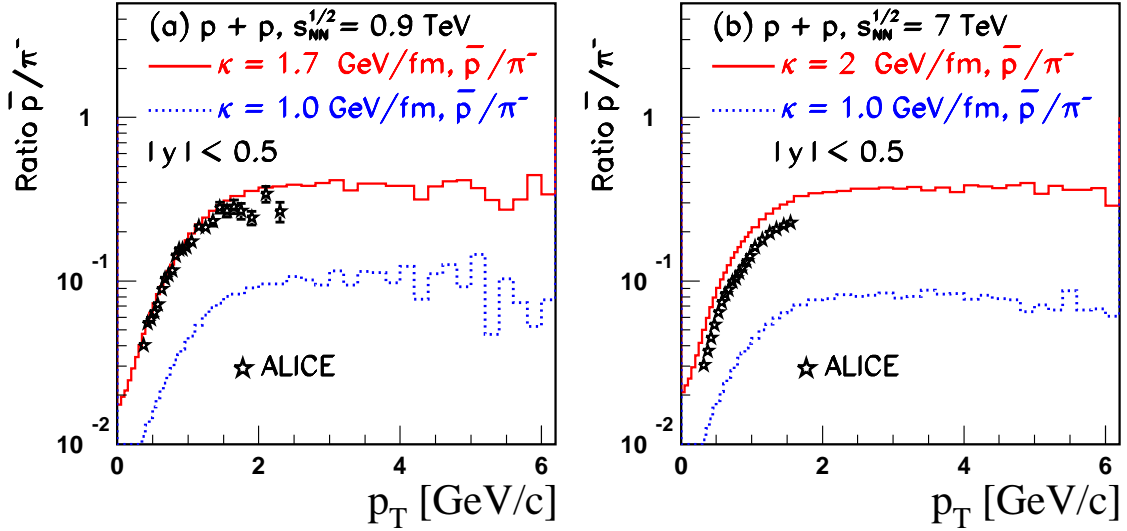}
\vskip 0.5cm\caption[pbar/pi- ratio at 0.9 Tev and 7 TeV] 
{\small (Color online) 
Comparison of {\small HIJING/B\=B v2.0} predictions 
with data on the
non-strange baryon over meson ratios from minimum bias events in the
rapidity range $|y| < 0.5$ at $\sqrt{s}$ = 0.9 TeV (left panel)
and at $\sqrt{s}$ = 7 TeV (right panel).
The solid and dashed lines have the same meaning as in
Fig.~\ref{fig:fig1}. Experimental results at 0.9 TeV 
are from Ref.~\cite{Aamodt:2011zj} (ALICE) and at 7 TeV from 
Ref.~\cite{Chojnacki:2011pv}(ALICE preliminary). 
The ratios have been calculated by us, dividing 
the spectra from Fig.~\ref{fig:fig3}.   
Error bars include only the statistical uncertainties.  
\label{fig:fig5}
}
\end{figure}

Figure~\ref{fig:fig5} shows a comparison of model 
predictions with ALICE data 
~\cite{Aamodt:2011zj,Chojnacki:2011pv} 
of non-strange baryon over meson ratios ($\bar{p}/\pi^-$) 
at 0.9 TeV and 7 TeV.
The ratios have been calculated by  dividing the spectra 
reported in Refs.~\cite{Aamodt:2011zj,Chojnacki:2011pv}.
Within our phenomenology the measured ratios are reasonably described 
in a scenario with SCLF effects (solid histograms).
The larger string tension parameterization results 
in a predicted increase of the ratio $\bar{p}/\pi^-$ 
by a factor of $\approx$ 5 at 7 TeV. Note that the models PYTHIA
\cite{Sjostrand:2007gs,Sjostrand:2006za} 
and EPOS \cite{Werner:2009zz,Werner:2010zz} 
cannot reproduce the observed high baryon-to-meson ratios
(see Fig.~6 and Fig.~7 in Ref.~\cite{Hippolyte:2009xz}).


The strange-particle ratios could also reveal 
manifestations of new collective phenomena.
In the EPOS model such an increase is obtained if the production 
of a {\em mini-plasma} is
assumed in $p+p$ collisions \cite{Abreu:2007kv}, \cite{Werner:2010zz}.
If confirmed by future measurements,
these observables could open a perspective on new physics in
$pp$ interactions. 


\begin{figure} [h!]
\centering
\includegraphics[width=0.9\linewidth,height=7.0cm]{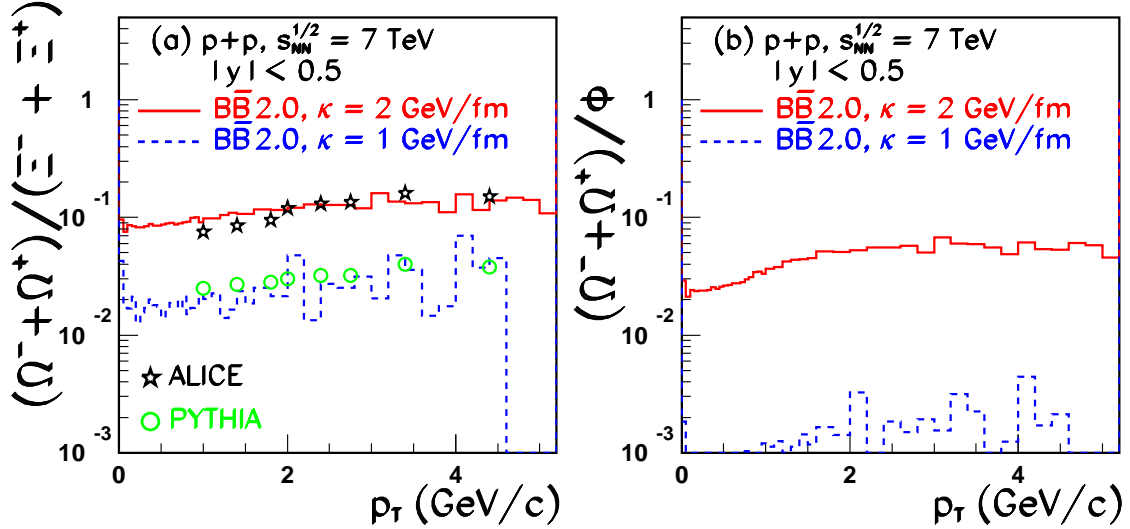}
\vskip 0.5cm\caption[strange baryon/meson ratio at 7 TEV -CMS data] 
{\small (Color online) {\small HIJING/B\=B v2.0}
  predictions of multi-strange hyperon ratios 
in $p+p$ collisions at $\sqrt{s}$ = 7 TeV for ratio $\Omega/\Xi$
(left panel) are compared to data. 
The calculated ratios (solid histogram) include the
combined effects of SCF and J\=J loops.
The dashed histogram are the results without SCF effects.
PYTHIA simulation results 
from Ref.~\cite{Chinellato:2011yn} are also included (left panel).
The data are from Ref.~\cite{Abelev:2012jp} (ALICE). 
Statistical error bars on the data points are smaller than the markers.
The model predictions for the ratio of multi-strange hyperons to mesons
$\Omega/\phi$ (right panel).    
\label{fig:fig6}
}
\end{figure}

To investigate possible differences in the production mechanisms of 
multi-strange baryons that do or do not contain a non-strange quark
we study the ratio of $\Omega^{-} (sss)$-to-$\Xi^{-} (dss)$ 
as a function of transverse momentum, $p_T$.
Due to low statistics we will consider the combined ratio
($\Omega^{-} + \overline{\Omega}^{+}$)-to-($\Xi^{-} +\overline{\Xi}^{+}$) 
at 7 TeV: the predictions are shown in Fig.~\ref{fig:fig6}a.
The model results without SLCF effects (dashed histogram)
underestimate the data by a factor of approximately 3.
Note that recent calculations with the {\small PYTHIA} event generator 
\cite{Sjostrand:2006za,Skands:2010ak,Field:2010bc} 
strongly underestimate the production rates \cite{Khachatryan:2011tm}
and fail to describe the above ratio ~\cite{Chinellato:2011yn}.
In contrast, the {\small HIJING/B\=B v2.0} model
in a scenario with SLCF effects 
(solid histograms) does describes the data.
This ratio could also help to study the possible 
saturation of the s-quarks, which would be indicated by a
flattening at high $p_T$. However, 
the currently available data do not allow for firm conclusions.

It has been argued that strangeness production
could be suppressed in $p+p$ collisions by the limited volume of 
the colliding system, which requires localized strangeness conservations
\cite{Kraus:2010bc}. Such a canonical suppression does not, however, explain
the suppression of the $\phi$ meson production in $p+p$ collisions 
because $\phi$ has a net strangeness of zero 
\cite{Becattini:2009zz,Xu:2008zzd}.
The study of the $\Omega/\phi$ ratio is therefore also of great
interest to distinguish between possible dynamical production mechanisms.  
The {\small HIJING/B\=B v2.0} model predictions are presented in 
Fig.~\ref{fig:fig6}b. The results predict a strong increase 
(up to an order of magnitude) for the scenario 
that includes SLCF effects (solid histogram).

In our approach, the dynamical
mechanism that leads to such high values of baryon-to-meson ratios   
is SLCF appearing at the initial stage of the interaction.
The SLCF mechanism strongly modifies the fragmentation processes
(strangeness suppression factors) and thus results in a large increase of
(strange)baryons. This interpretation is also supported by more sophisticated
theoretical calculations, in a scenario in which a 
time-dependent pulse for the initial strength of
the color field is considered \cite{Skokov:2009zz}.
An observed large enhancement of the baryon-to-meson ratios 
would be consistent with SLCF playing
an important role in multiparticle production in $p+p$
collisions at LHC energies and suggesting that high energy density fluctuations
can reach very high densities, potentially comparable to those reached 
in central Au + Au collisions at RHIC energies \cite{Alver:2010ck}.



\subsection{Baryon-to-meson ratio in HM pp collision events}

To test the above assumptions, and
in order to study possible new phenomena in $p+p$ collisions, 
we examine in this paper the dependence of particle production 
as a function of the total charged particle multiplicity ($N_{\rm ch}$)
and compare the results with those from minimum bias event selection.
The MB event selection is defined here as 
the existence of one charged particle in the rapidity interval $|y| < 1$. 
Since (multi)strange particle production in heavy-ion reactions is 
enhanced relative to that in MB $p+p$ collisions~\cite{ToporPop:2011wk}, 
one might ask whether the production rates in HM
$p+p$ collisions may already exhibit any feature like an enhancement 
due to SLCF effects. 
The increase is quantified by calculating the $p_T$ dependence 
at mid-rapidity ($|y|< 0.8$) of the baryon/meson yield ratios, 
{\it e.g.}, (multi)strange baryon ($\Lambda$, $\Xi$, and $\Omega$) over mesons
(pions).

\begin{figure} [h!]
\centering
\includegraphics[width=0.9\linewidth,height=13.5cm]{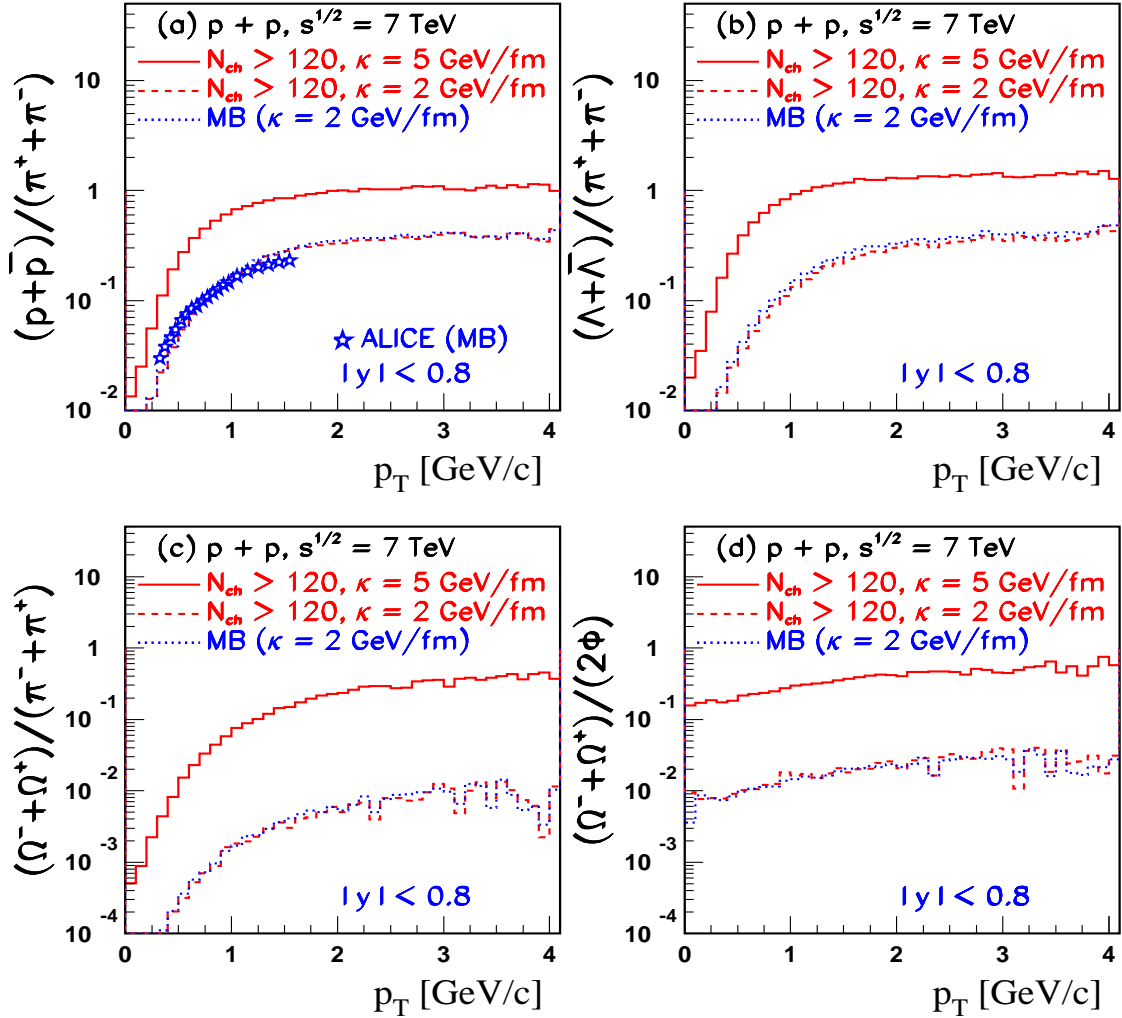}
\vskip 0.5cm\caption[strange baryon/meson ratio at 7 TEV -CMS data] 
{\small (Color online) 
The {\small HIJING/B\=B} v2.0 model predictions for production ratios 
[$(p+\bar{p})/(\pi^+ + \pi^-)$, (a)], 
[$(\Lambda+\bar{\Lambda})/(\pi^+ + \pi^-)$, (b)] 
[$(\Omega^-+\Omega^+)/(\pi^+ + \pi^-)$, (c)]
[$(\Omega^-+\Omega^+)/ 2 \phi$, (d)].
The results are plotted at mid-rapidity ($|y|<0.8$) taking 
$\kappa = 2$ GeV/fm in the MB event selection (dotted histograms) and  
HM events with $N_{\rm ch} > 120$ (dashed histograms).
For the HM events selection 
the predictions are also shown considering 
$\kappa = 5$ GeV/fm (solid histograms).
The experimental ratios in (a) have been calculated 
by dividing the spectra from Ref.~\cite{Chojnacki:2011pv}.
\label{fig:fig7}
}
\end{figure}


Figure~\ref{fig:fig7} shows the model predictions for various 
baryon-to-meson ratios as a function of transverse momentum. Shown are the 
results for non-strange baryon over non-strange 
meson production [$(p+\bar{p})/(\pi^+ + \pi^-)$, part (a)], 
strange baryon over non-strange
meson production [$(\Lambda+\bar{\Lambda})/(\pi^+ + \pi^-)$, part (b)] 
multi-strange baryon
over non-strange meson production 
[$(\Omega^-+\Omega^+)/(\pi^+ + \pi^-)$, part (c)],
and multi-strange baryon over $\phi$ mesons 
[$(\Omega^-+\Omega^+)/ 2 \phi$, part (d)] for MB and 
HM ($N_{\rm ch} > 120$) events.
The results are presented for two scenarios:
MB and HM ($N_{\rm ch} > 120$) events assuming   
$\kappa$= 2 GeV/fm,
and for HM events assuming $\kappa = 5$ GeV/fm.

One first notes that, for a constant value of $\kappa$, there are
negligible differences between the predicted ratios for MB and HM 
events. On the contrary, assuming a larger value of $\kappa$ leads
to large increases that depend on particle species. 
The ratio of non-strange baryon over non-strange mesons 
$(p+\bar{p})/(\pi^+ + \pi^-)$) shows a multiplicative enhancement
of approximately
$3$ relative to MB and HM results with $\kappa$= 2 GeV/fm
(Fig~\ref{fig:fig7} a).
However, the enhancement for (multi)strange baryon over meson ratio
increases with increasing mass of hyperons: up to a factor of four
 for $(\Lambda+\bar{\Lambda})/(\pi^+ + \pi^-)$ (Fig.~\ref{fig:fig7} b),
up to a factor of ten
for $(\Xi^-+\Xi^+)/(\pi^+ + \pi^-)$ (not shown here),
and up to a factor of twenty
for $(\Omega^-+\Omega^+)/(\pi^+ + \pi^-)$ (Fig.~\ref{fig:fig7} c)
and $(\Omega^-+\Omega^+)/ 2 \phi$ (Fig.~\ref{fig:fig7} d).

Recently, new (preliminary) data for the ratio of non-strange baryons 
over mesons, $(p+\bar{p})/(\pi^+ + \pi^-)$ as function of charged particle
multiplicity have been reported by the CMS 
collaboration\cite{cms_id_march2012}.
These data do not show a dependence of the measured ratio on the 
multiplicity of the event, consistent with a scenario assuming 
a constant value for $\kappa$.
However, the measurements are performed within a very 
limited $p_T$ range ($p_T < 1.2 $ GeV/c) and the selection of events 
is neither NSD nor INEL and make comparison with theory difficult. 
Data on (multi)strange particle production extending over a larger 
$p_T$ range are more sensitive to SCF and would allow drawing 
a more definitive conclusion.

With high statistics measurements 
of identified particles, the LHC collaborations 
could test the model predictions and lend credence to the idea   
that new phenomena or possible out of equilibrium  
mQGP has been formed in HM $p+p$ collisions.
Note that the ALICE collaboration recently reported measurements of 
the inclusive $J/\psi$ yield as a function of charged particle
density at mid-rapidity ($|\eta|< 1$) in MB and HM events at 7 TeV.
HM events were selected with different bins, up to four times 
MB multiplicity density.
In these HM events
an enhancement by a factor of about eight for $J/\psi$ yields
at mid-rapidity ($|y| < 0.9$) 
was found relative to those in MB events~\cite{Abelev:2012rz}.

Preliminary ALICE data~\cite{Schuchman2012}
on $\Lambda$ and $K_S^0$ nuclear modification factors in central 0-5\%
Pb+Pb collisions at 2.76A TeV appear also to agree qualitatively 
with our predictions of enhanced
hyperon/meson yield ratios and their nuclear quenching pattern
at transverse momentum $p_T > 2 $ GeV/{\it c} 
(see Fig. 8 from Ref.~\cite{ToporPop:2011wk} ). 
In this paper we propose that
a similar enhancement of the baryon-to-meson ratios 
may be observed  in rare HM $p+p$ collisions.  
Just as in the Pb+Pb collisions, the hyperon/meson enhancement 
and the quenching pattern of $\Lambda$, $K_S^0$ and $\pi^{\pm}$
in HM $p+p$ collision events relative to those in MB,
corroborated with other bulk flow correlations, could provide further 
evidence of out of equilibrium mQGP production in rare HM $p+p$ reactions.

A mini quark-gluon plasma differs from the  
strong coupling plasma (sQGP) produced in central 
nucleus-nucleus ($A+A$) collisions, 
mostly by its small initial transverse size $R_p \sim 1$ fm compared to 
the significantly larger nuclear transverse size $R_{A} \sim 5$ fm.
Extensive hydrodynamic~\cite{Kolb:2003dz} 
and transport calculations~\cite{Molnar:2001nk}
have shown that collective flow signatures, such as perfect fluid 
elliptic flow, require several fm/{\it c} to develop. 
Similarly, jet quenching observables depend strongly on the jet path
length and thus, in $p+p$ collisions quenching will be much weaker
than in Pb+Pb central collisions.
In contrast, signatures associated with strangeness 
equilibration~\cite{ToporPop:2011wk},\cite{Muller:2011tu}
are generated on much faster time scales. Therefore, hyperon-to-meson 
ratios and their $p_T$ dependence may serve as 
the best probe of possible out of equilibrium mQGP formation.


In order to explore better the enhancement of the 
baryon/meson production ratio 
as a possible signature of {mQGP} in $p+p$ collisions, and 
especially its possible dependence on event multiplicity, 
the particle yields at  mid-rapidity ($|y| < 0.8$)
are studied in six bins of charged particle multiplicity: 
$10 \leq N_{\rm ch} < 30$; $30 \leq N_{\rm ch} < 60$; 
$60 \leq N_{\rm ch} < 80$; $80 \leq N_{\rm ch} < 100$; 
$100\leq N_{\rm ch} < 120$; $N_{\rm ch} \geq 120$.

We select an experimental observable that would be sensitive to 
possible new phenomena and is well adapted to the 
low statistics expected from low yield particle production.
We consider the integrated values $Y_\kappa$ 
of the baryon-to-meson yield ratios shown in Fig.~\ref{fig:fig7}.
The domain of the integration is taken over the range 
$1 < p_T < 4$ GeV/{\it c}, where the increase appears to be strongest.
The calculations of $Y_\kappa$ are performed for the above six bins of 
multiplicity. We also define $Y_{\rm MB}$ for MB events, 
{\it i.e.}, the integrated ratio without selection on multiplicity.
 For  $Y_{\rm MB}$ the mean value of the string tension is taken as 
$\kappa=\kappa(s) \approx 2$ GeV/fm. For the multiplicity bins,
the calculations are performed for both 
$\kappa=\kappa(s) \approx 2$ GeV/fm ($Y_2$) and
$\kappa = 5$ GeV/fm ($Y_5$).
Since we expect a gradual transition as function of multiplicity an 
intermediate value $\kappa = 3$ GeV/fm ($Y_3$) is also considered.

\begin{figure} [h!]
\centering
\includegraphics[width=0.9\linewidth,height=7.0cm]{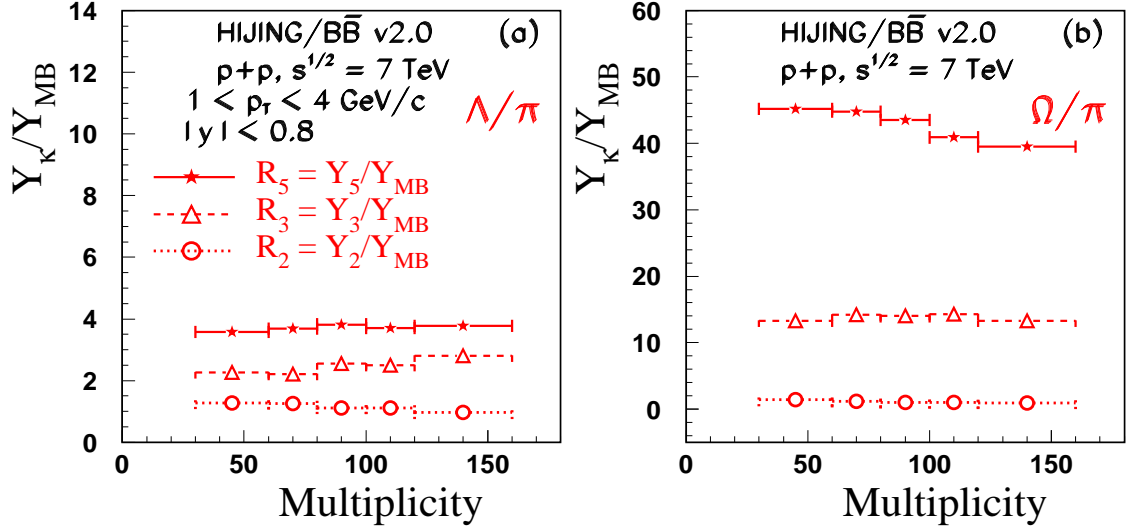}
\vskip 0.5cm\caption[strange baryon/mesons ratio hm events] 
{\small (Color online) {\small HIJING/B\=B v2.0}
 predictions for the relative increase  
 $R_\kappa=Y_\kappa/Y_{\rm MB}$ (see text for explanation). 
The results are plotted 
as a function of multiplicity in five bins: 
$30 \leq N_{\rm ch} < 60$; $60 \leq N_{\rm ch} < 80$;
$80\leq N_{\rm ch} < 100$; $100 \leq N_{\rm ch} < 120$; $N_{\rm ch} > 120$.
The values $R_\kappa$ are given for  
$(\Lambda+\bar{\Lambda})/(\pi^+ + \pi^-)$ (left panel) and for 
 $(\Omega^-+\Omega^+)/(\pi^+ + \pi^-)$ (right panel).
The results for the relative increase 
$R_3 = Y_3/Y_{\rm MB}$ (triangles) 
and $R_5 = Y_5/ Y_{\rm MB}$ (stars)
are obtained using $\kappa = 3$ GeV/fm and 
$\kappa = 5$ GeV/fm, respectively. The integrated values 
$Y_{\rm MB}$ are obtained using 
$\kappa = \kappa(s) \approx 3$ GeV/fm~\cite{ToporPop:2010qz}. 
\label{fig:fig8}
}
\end{figure}

In Figure~\ref{fig:fig8},  
the theoretical predictions for the relative increase in the 
integrated values $Y_{\kappa}$ to the integrated values $Y_{\rm MB}$,
{\it i.e.}, the ratio $R_{\kappa} = Y_{\kappa}/Y_{\rm MB}$, are shown.
Note that the statistical fluctuations are too large in the first bin
($10 \leq N_{\rm ch}<30$) so it is not included in the plot.
The calculations are presented for the ratios 
$R_2$= $Y_{2}/Y_{\rm MB}$, $R_3$= $Y_{3}/Y_{\rm MB}$, and 
$R_{5}= Y_5/Y_{\rm MB}$. 

As examples, the predictions for (multi)strange-baryon-to-meson ratios
are given, {\it i.e.}, $(\Lambda+\bar{\Lambda})/(\pi^+ + \pi^-)$ 
and  $(\Omega^-+\Omega^+)/(\pi^+ + \pi^-)$.
For the $(\Lambda+\bar{\Lambda})/(\pi^+ + \pi^-)$ ratio,
the model predicts an almost constant value of $R_{\kappa}$
with no multiplicity dependence,
while a slight dependence is predicted for the ratio $R_5$ of 
$(\Omega^-+\Omega^+)/(\pi^+ + \pi^-)$.
For the ratio $(\Lambda+\bar{\Lambda})/(\pi^+ + \pi^-)$
an enhancement of a factor of 4 of $R_5$ over $R_2$ is predicted
(Fig.~\ref{fig:fig8}a).
 In contrast, for  $(\Omega^-+\Omega^+)/(\pi^+ + \pi^-)$
(Fig.~\ref{fig:fig8}b)
higher values (up to a factor of approximately 40)
are predicted in the scenario with a possible 
transition to an out of equilibrium mQGP state, 
$R_5$ (solid line), than in a scenario without, $R_2$ (dotted line).

If the assumption of the 
dependence of $\kappa$ on the total charged particle
multiplicity ($N_{\rm ch}$) is valid,
we expect to see in the data a transition from $R_2$ (dotted line)
at low multiplicity to  $R_3$ (dashed line),
to $R_5$ values (solid line) for higher multiplicity events.
These results strongly suggest that 
the experimental data could contain a signature for a 
possible transition to a mQGP phase in hadronic collisions. 
The shape of an observed transition 
as a function of multiplicity could 
contain information on the nature of the underlying physics,
{\it e.g.}, if the transition is smooth or has a net threshold. 

We showed that the baryon-to-meson ratios  
have an enhancement up to the highest LHC energy (14 TeV)
~\cite{ToporPop:2010qz}.
Based on Eq.~\ref{eq:eq4} the string tension value
has a predicted modest increase from $\approx 2$ GeV/fm to 
$\approx 2.15 $ GeV/fm, when going from  
$\sqrt{s} = 7$ TeV to  $\sqrt{s} = 14$ TeV.
In addition, as shown in Ref.~\cite{ToporPop:2010qz}, 
a saturation sets in near  $\sqrt{s} \approx 3$ TeV.
Therefore, we expect relatively small further increases in
the strange particle ratios and in the predicted values
for $R_{\kappa}$ in HM $p+p$ collisions at higher LHC energy.

\section{Conclusions}

In this work within the phenomenology of the 
{\small HIJING/B\=B} v2.0 model we discussed observables sensitive
to possible new phenomena, such as strong longitudinal color fields
in HM $p+p$ collision events.
For MB bias events we show that a good description is obtained for charged 
and identified particle production, 
taking an energy dependence of mean string tension values 
$\kappa = \kappa(s)= \kappa_{0} (s/s_{0})^{0.04}\,\,{\rm GeV/fm}$.

The predictions for baryon/meson production ratios 
in $p+p$ collisions at $\sqrt{s} = 7$ TeV are discussed.  
We analyze the dependence of these ratios on the degree of collectivity
in the reaction dynamics,
characterized by an infrared sensitive variable, 
the string tension $\kappa$.
The formation of a collective phase in high multiplicity
$p+p$ collisions would be made evident within our phenomenology
by a relative increase in the baryon/meson ratios with 
increasing multiplicity,
particularly those ratios involving strange and multi-strange particles.

The experimental data could show 
a multiplicity dependent transition as indicated by comparing 
results obtained with a lower value of 
$\kappa = 2$ GeV/fm to those corresponding to a higher value of 
$\kappa = 5$ GeV/fm originating from possible production of 
a transitory out of equilibrium mQGP phase.
This transition is most likely gradual as a function of multiplicity 
and center-of-mass energy $\sqrt{s}$ .
Corroborated with other observables sensitive to collective behavior,
such as {\em ridge structure}, enhanced radial flow, high $p_T$ particle
suppression ({\em jet quenching}) in $p+p$ collisions, 
observation of saturation of hyperon/meson ratios
could provide possible signatures of 
out of equilibrium  mQGP phase 
formed in high multiplicity $p+p$ collision events at ultra-high energies.

\section{Acknowledgments}
\vskip 0.2cm 

{\bf Acknowledgments:} We thank S. Das Gupta for useful 
discussions and continue support.   
VTP acknowledges discussion with M. Petrovici and 
access to computer facilities IFIN-HH, Bucharest, Romania and at Columbia
University, New York, where parts of these calculations were performed.
This work was supported by the Natural Sciences and Engineering 
Research Council of Canada.  
This work was also supported by the Division of Nuclear Science, 
of the U. S. Department of Energy under Contract No. DE-AC03-76SF00098 and
DE-FG02-93ER-40764.

\end{document}